\def \abc#1#2#3#4 {\reference#1, {\sl#2}, {\bf#3}, #4}
\def \blank {\lower 5pt\hbox to 0.75in{\hrulefill}}

\def \cm{~\rm{cm}}
\def \s{~\rm{s}}
\def \km{~\rm{km}}

\def \AU{~\rm{AU}}

\def \yr{~\rm{yr}}

\def \erg{~\rm{erg}}

\def \lae{\mathrel{<\kern-1.0em\lower0.9ex\hbox{$\sim$}}}
\def \gae{\mathrel{>\kern-1.0em\lower0.9ex\hbox{$\sim$}}}

\documentclass[12pt,preprint]{aastex}
\begin{document}

\title{FORMATION OF PLANETARY NEBULA LOBES BY JETS}

\shorttitle{Arcs in AGB}
\shortauthors{Soker}

\author{Noam Soker\altaffilmark{1}}
\affil{University of Virginia, Department of Astronomy, P.O.~Box 3818, 
       Charlottesville, VA 22903-0818, USA}
\email{soker@physics.technion.ac.il}

\altaffiltext{1}{On Sabbatical from the University of Haifa at Oranim,
       Department of Physics, Oranim, Tivon 36006, Israel.} 

\begin{abstract}

 I conduct an analytical study of the interaction of jets, or a 
collimated fast wind (CFW), with a previously blown asymptotic giant branch
(AGB) slow wind. Such jets (or CFWs) are supposedly formed when a compact 
companion, a main sequence star or a white dwarf, accretes mass from 
the AGB star, forms an accretion disk, and blows two jets. 
 This type of flow, which is thought to shape bipolar planetary nebulae 
(PNe), requires 3-dimensional gas dynamical simulations, which are 
limited in the parameter space they can cover. 
By imposing several simplifying 
assumptions, I derive simple expressions which reproduce some basic 
properties of lobes in bipolar PNe, and which can be used to guide future 
numerical simulations. I quantitatively apply the results to two proto-PNe. 
I show that the jet interaction with the slow wind can form lobes which 
are narrow close to, and far away from, the central binary system, and
which are wider somewhere in between.  Jets that are recollimated and have 
constant cross section can form cylindrical lobes with constant diameter, 
as observed in several bipolar PNe.  Close to their source, jets blown by 
main sequence companions are radiative; only further out they become 
adiabatic, i.e., they form high-temperature low-density bubbles that inflate 
the lobes. This implies that radiative cooling must be incorporated in 
numerical codes intended to study the formation of lobes in PNe.
  
\end{abstract}


\clearpage 

\section{INTRODUCTION}

 Planetary nebulae (PNe) and proto-PNe possess a rich spectrum of 
different structures, with spherical and bipolar PNe at the two extremes
of that spectrum.  
 Bipolar PNe are defined as extreme aspherical PNe
having two lobes with an `equatorial' waist between them
(Schwarz, Corradi \& Stanghellini 1992).
 Several models have been suggested to explain the formation of lobes.
 One of the popular models a decade ago was the interacting wind model,
where a fast wind blown by the central star of the PN
is collimated by a previously ejected AGB stellar dense equatorial wind
(e.g., Balick 1987; Soker \& Livio 1989; Frank \& Mellema 1994; 
Mellema \& Frank 1995; Mellema 1995).
 When the equatorial to polar density ratio is very high, a bipolar nebula
was supposed to be formed.
However, the numerical simulations did not form bipolar PNe with
very narrow waists, but rather formed elliptical PNe, or bipolar PNe
with wide waists.
 Another problem of the interacting wind model is the high momentum
and kinetic energy observed in several bipolar PNe and proto-PNe
(Bujarrabal {\it et al.} 2001; see Balick 2000 for some other problems).
  Therefore, the simple interacting wind model cannot explain many of the
observed bipolar PNe.
 It seems that in order to form many bipolar PNe, e.g., those
with very narrow waists, a collimated fast wind (CFW) is required
(Morris 1987; Soker \& Rappaport 2000, hereafter SR00).
 Such preliminary simulations, but where the CFW was blown by the primary
star, were performed by Frank, Ryu, \& Davidson (1998),
who indeed got very narrow waists in their simulated nebulae.
 Other types of models are based on magnetic fields playing a dynamical
role in the mass loss process from AGB stars (e.g.,  Garc\'ia-Segura 1997; 
Garc\'ia-Segura \& L\'opez 2000; Blackman {\it et al.} 2001).
 There are some severe problems with dynamical magnetic models
(Soker \& Zoabi 2002). 
 The main problem is that the AGB has to be spun-up by a stellar companion
in order to possess the required activity; hence we
are in the regime of binary models. 
 
 As reviewed recently by SR00, the most promising mechanism for the 
formation of bipolar PNe with narrow waists is based on an accreting 
compact companion to the mass losing AGB star.
 In most cases, the companion is a white dwarf or a main sequence star
(SR00). 
 When a compact binary companion accretes matter with sufficiently 
high specific angular momentum, an accretion disk is formed. 
When the accretion rate is above some threshold (SR00), two jets 
are expected to be blown by the companion. 
 If the jets are not well collimated, the outflow from the companion 
is termed CFW. 
 In addition, the companion may lead to a concentrated equatorial flow
(Mastrodemos \& Morris 1999), further increasing the equatorial to
polar density ratio.
 The formation of bubbles (or lobes) by jets have been studied in 
a variety of astrophysical flows, e.g, young stellar objects
(e.g., Masson \& Chernin 1993), clusters of galaxies (e.g., Reynolds, 
Heinz \&  Begelman 2001), and gamma ray burst environments 
(Ayal \& Piran  2001). 
 However, the condition of jets blown into AGB winds are different
from those in these cases. 

 To study the interaction of the two winds blown simultaneously, the slow
wind by the AGB star and the CFW (or two jets) by the accreting
compact companion (see figs. 1 and 2 in SR00), a 3D gasdynamical
numerical code is required.  
 Such numerical simulations require heavy computer resources,  
hence can cover only a small number of cases. 
 Therefore, it will be helpful to derive simple expressions than can both
describe the general structure of lobes obtained by different jets,
and guide future numerical simulations. 
 This paper is aimed at reaching these objectives, by exploring part of 
the parameter space of jets interacting with spherical AGB winds,
and under simplifying assumptions. 
  To manifest analytical solutions, I consider two extreme cases,
which are described in $\S 2$, and are studied in $\S 3$ and $\S 4$. 
A short summary is in $\S 5$.

\section{JET PROPAGATION}

 In the present analytical study, I treat the flow at large distances from
the binary system, neglecting the deflection of the jets (or CFW; see SR00).
 I assume an axisymmetric flow in which a jet expands along the 
symmetry axis and into a previously ejected spherically symmetric 
slow AGB wind.
 The slow wind density as a function of distance from the binary system
is $\rho_s=\dot M_s/(4 \pi v_s r^2)$, where $v_s$ is the slow wind speed,
and $\dot M_s$ is the mass loss rate, both are constant, and 
where mass loss rate is defined positively. 
The gas inside the undisturbed jets, one at each side of the equatorial 
plane, expands at a constant speed $v_j$, the mass loss rate into each 
jet is $\dot M_j$, and each jet expands into
a solid angle of $\Omega_j = 4 \pi \beta$.  
 The density inside an undisturbed radially expanding jet, i.e.,
its cross section increases as $r^2$, is 
$\rho_j=\dot M_j/(4 \pi \beta v_j r^2)$.
 The velocity of the jet head, $v_h$, is determined by equating the
slow wind pressure on the jet head with that of the jet material.
 Since thermal pressure can be neglected in the pre-shocked media,
the expression is $\rho_j (v_j-v_h)^2 = \rho_s (v_h-v_s)^2$.
 In the present analytical study I treat two extreme cases: 
slow-propagating and fast-propagating jets.  
 In the slow-propagating jet case the jet head proceeds at a speed
$v_h \ll v_j$, while in the fast-propagating jet case the jet head 
does not slow much, $v_s \ll v_h \lesssim v_j$.  

 The velocity of a slow-propagating jet to be treated in the next section
is given by 
\begin{eqnarray} 
v_h \simeq v_s \left[ 1 + 
\left( \frac{\dot M_j v_j }{\beta \dot M_s v_s} \right)^{1/2} \right]; 
\qquad {\rm for} \qquad \rho_s \gg \rho_j.   
\end{eqnarray}
 I therefore assume that 
 $\dot M_j v_j \lesssim {\rm few} \times 10 \beta \dot M_s v_s$,
so that $v_h \sim 2-10 v_s$. 
 Since I neglect the bending of the jet by the AGB slow wind
close to the binary system, the momentum flux of the jet must
be (SR00) $\dot M_j v_j \gtrsim \dot M_s v_s \tan \alpha$,
where $\alpha$ is the half opening angle of the jet (SR00). 
 For $\tan \alpha \ll 1$, $\tan \alpha \simeq 2 (\beta)^{1/2}$.
 For $\alpha =10^{\circ}$, for example, $\beta = 0.0076$, 
$2 \beta^{1/2} = 0.174$, and $\tan \alpha =0.176$. 
Over all, the treatment of the slow-propagating jet is 
applicable to the case 
$(\dot M_j v_j)/(\dot M_s v_s) \sim 0.1-0.5$, for a half opening
angle of $\alpha \sim 5-30 ^\circ$.    
 However, since, as we'll see later, the derived expressions are not 
so sensitive to the different variables, the results will be 
applicable to a much larger parameter space.   
  
 From the expressions developed above, we can find the density of
the jet shocked material, and from that, the radiative cooling time,
$t_{\rm cool} = (5/2) n kT /(n_e n_p \Lambda)$, where
 $n$, $n_e$, and $n_p$, are the total number, electron, and proton, 
post-shock densities, respectively, 
taken to be four times the pre-shock densities, 
and for the considered temperature range, which is determined
by the speed of the jet material,
$\Lambda \simeq 10^{-22} \erg \cm^3 \s^{-1}$. 
 The cooling time of the shocked jet material is obtained by substituting
typical values
\begin{eqnarray} 
t_{\rm cool} ({\rm jet}) \simeq 17 
\left( \frac {v_j} {400 \km \s^{-1}} \right)^3 
\left( \frac {\dot M_j} {10^{-7} M_\odot \yr ^{-1}} \right) ^{-1} 
\left( \frac {\beta} {0.01} \right)   
\left( \frac {r} {10^{16} \cm} \right) ^{2} \yr.  
\end{eqnarray}
 This time scale is much longer than the flow time along the
jet diameter. 
 However, the cooling time of the post shocked material should 
be treated differently and compared with another flow time scale.
This is done for the two flow cases in the next two sections.
 
\section{SLOW-PROPAGATING JETS}
\subsection{Jet Interaction with the Slow Wind}
  The shocked jet material will expand and form a hot low density
 bubble that will accelerate the dense slow wind's material around it.
  The results of Castor, McCray \& Weaver (1975) for a stellar wind 
bubble in a dense interstellar medium can be used. 
 Using their equation (6) for the radius of the expanding bubble as a 
function of the expansion time of the bubble, $t_b$, gives 
\begin{eqnarray} 
R_b =  0.76 \left( \frac{0.5 \dot M_j v_j^2}{\rho_s} \right)^{1/5} t^{3/5}=
1.0 \times 10^{15} 
\left( \frac {v_j} {400 \km \s^{-1}} \right)^{2/5} 
\left( \frac {\dot M_j} {10^{-7} M_\odot \yr ^{1/5}} \right) ^{1/5}
\nonumber \\ \times
\left( \frac {v_s} {15 \km \s^{-1}} \right)^{1/5} 
\left( \frac {\dot M_s} {10^{-5} M_\odot \yr ^{-1}} \right) ^{-1/5}  
\left( \frac {r} {10^{16} \cm} \right) ^{2/5}  
\left( \frac {t_b} {100 \yr } \right)^{3/5} \cm ,  
\end{eqnarray}
where the local slow wind density at $r$ was taken in the second equality.
 The expansion velocity of the bubble's surface is
\begin{eqnarray} 
v_b = 0.6 R_b/t = 
 20
\left( \frac {v_j}{400 \km \s^{-1}} \right)^{2/5} 
\left( \frac {\dot M_j} {10^{-7} M_\odot \yr ^{1/5}} \right) ^{1/5}
\left( \frac {v_s} {15 \km \s^{-1}} \right)^{1/5} 
\nonumber \\ \times
\left( \frac {\dot M_s} {10^{-5} M_\odot \yr ^{-1}} \right) ^{-1/5}  
\left( \frac {r} {10^{16} \cm} \right) ^{2/5} 
\left( \frac {t_b} {100 \yr } \right)^{-2/5} \km \s^{-1} .  
\end{eqnarray}
   
 The expansion time of the hot bubble, $t_b$, should be compared with the
cooling time of the hot gas inside the bubble.
The density inside the bubble is given by 
$\rho_b = \dot M_j t_b /(4 \pi R_b^3/3)$, from which the 
cooling time is derived 
\begin{eqnarray} 
t_{\rm cool}({\rm bubble}) \simeq 30
\left( \frac {v_j}{400 \km \s^{-1}} \right)^{2} 
\left( \frac {\dot M_j} {10^{-7} M_\odot \yr ^{-1}} \right) ^{-1} 
\left( \frac {R_b} {5 \times 10^{15} \cm} \right) ^{3}   
\left( \frac {t_b}{100 \yr} \right)^{-1} \yr. 
\end{eqnarray}
 The condition for the hot gas (the shocked jet material) to 
inflate a bubble is that the cooling time is longer than the expansion time,
i.e., $t_{\rm cool}({\rm bubble}) > t_b$. 
 Substituting for $t_{\rm cool} ( {\rm bubble})$ from equation (5),
and eliminating $R_b$ by equation (3), gives the radius outward to which
the shocked jet material will not have time to cool before 
inflating a large bubble (which will evolve to a nebular lobe) 
\begin{eqnarray} 
r_s \gtrsim 4.4 \times 10^{15}
\left( \frac {v_j} {400 \km \s^{-1}} \right)^{-8/3} 
\left( \frac {\dot M_j} {10^{-7} M_\odot \yr ^{-1}} \right) ^{1/3}
\left( \frac {v_s} {15 \km \s^{-1}} \right)^{-1/2} 
\nonumber \\ \times
\left( \frac {\dot M_s} {10^{-5} M_\odot \yr ^{-1}} \right) ^{1/2}  
\left( \frac {t_b} {100 \yr } \right)^{1/6} \cm .  
\end{eqnarray}

 The scaling in the last equation shows that the adiabatic phase, where
the cooling time of the shocked material is very long, starts
at $r \sim 10^{16} \cm$. 
 However, the parameters change from one system to another. 
For example, in the proto-bipolar PN OH 231.8+4.2,  the mass loss rate
is very high, $2 \times 10^{-4} M_\odot \yr^{-1}$
(Alcolea {\it et al.} 2002 and more references therein). 
 If we take the mass loss rate in both the AGB wind and the jet
to be 10 times higher than the values used in the scaling of equation (6),
we find the adiabatic phase to start at $r_s \sim 5 \times 10^{16} \cm$
from the central source. 
 If the companion accretes via a Roche lobe overflow, then the mass
loss rate into the jets can be much higher. 
 In OH 231.8+4.2, the mass loss rate into each of the two jets could
have been $\sim 10^{-4} M_\odot \yr ^{-1}$ for a short time during
a Roche lobe overflow (Soker 2002). 
 If we assume that the opening angles of the two jets were large,
so the jets' density is lower than the slow wind density, we can
use equation (6). 
 Substituting $\dot M_s =10^{-4} M_\odot \yr^{-1}$, 
$\dot M_j=10^{-4} M_\odot \yr^{-1}$, and $v_j =500 \km \s^{-1}$,
we find from equation (6) $r_s \simeq 5 \times 10^{16} \cm$.
  This is a substantial fraction of the lobe size 
(Kastner \& Weintraub 1995).
 Therefore, during a large fraction of the evolution (since the jet phase 
was short; Soker 2002) the flow was momentum conserving (radiative), i.e.,
the shocked material was cooling on a relatively short time. 
  Indeed, Bujarrabal {\it et al.} (2002), ruled out the adiabatic case for 
OH 231.8+4.2, mainly on the ground of the shape of the lobes, which are 
not inflated as is expected in the adiabatic flow.
 
\subsection{The Bubble Shape}
  From equation (3) it turns out that the radius of a spherical bubble
 increases with $r$ much slower than $r$ does.
 For example, at a distance from the center 32 times that was is
 used in equation (3) the bubble radius will be only 4 times larger,
keeping the time and the other parameters unchanged. 
 However, the bubble will not be spherical since the jet propagates
outward, and a more accurate treatment is required. 
I assume that the bubble expands perpendicular to the symmetry axis,
along the $z$ direction, i.e., parallel to the equatorial plane. 
 I also take the density in planes parallel to the equatorial plane
to be the same as the density where the plane intersects the
symmetry axis. The distance of the plane from
the equatorial plane is the coordinate $x$. 
 The relevant quantity is the energy deposited per unit
length along the jets propagation, which is 
$e_0 = 0.5 \dot M_j v_j^2 / v_h$. 
 I use the relation between pressure and energy and the momentum equations
as in Castor {\it et al.} (1975), and neglect the thermal energy
of the swept slow wind material. 
 The last assumption is adequate since I neglect the slow wind radial
expansion and the dependence of density on the distance from 
the symmetry axis,
while the assumptions allow a simple expression for the energy equation. 
 The energy equation reads $e = e_0-e_k$, where $e$ is the thermal 
energy per unit length (along the symmetry axis) in the bubble, and 
$e_k$ is the kinetic energy per unit length of the swept up slow wind gas. 
 Neglecting the density variation perpendicular to the symmetry axis and
the radial expansion of the slow wind gives the distance of the 
bubble surface from the symmetry axis
\begin{eqnarray}
z_b (x)= \left( \frac{16}{5 \pi} \frac {e_0}{\rho_s(x)}\right)^{1/4} t^{1/2}
\simeq \left( \frac {e_0}{\rho_s(x)}\right)^{1/4} t^{1/2},   
\end{eqnarray}
where $x$ and $z$ are the coordinates parallel and perpendicular to the
symmetry axis, respectively. 
 Substituting in the typical values used here, with $v_h$ by equation (1),
gives 
\begin{eqnarray} 
\frac {z_b}{x} \simeq 1.3
\left( \frac {v_j} {400 \km \s^{-1}} \right)^{1/2} 
\left( \frac {\dot M_j} {10^{-7} M_\odot \yr ^{-1}} \right) ^{1/4}
\left( \frac {v_h} {4 v_s} \right)^{-1/4} 
\left( \frac {\dot M_s} {10^{-5} M_\odot \yr ^{-1}} \right) ^{-1/4}  
\nonumber \\ \times
\left[ \frac {t_b(x)} {100 \yr } \right]^{1/2} 
\left( \frac {x} {10^{16} \cm } \right)^{-1/2}.
\end{eqnarray}
 For the same parameters, the velocity is 
\begin{eqnarray} 
v_b \equiv \frac {d z_b}{dt}  \simeq 0.5 \frac{z_b}{t} \simeq 20  
\left[ \frac {t_b(x)} {100 \yr } \right]^{-1/2} 
\left( \frac {x} {10^{16} \cm } \right)^{1/2} \km \s^{-1}.
\end{eqnarray}
 Because of the assumption that density does not depend on the $z$ 
coordinate, the last two equations are less accurate for $z_b > r$, 
i.e., close to the central binary system. 
 
 It should be noted that in the last two equations the time $t_b(x)$ is 
counted from the moment the jet reaches the point $x$, hence $t_b$ is 
longer for lower values of $x$, i.e., closer to the center of the 
nebula.
 If the nebula is observed at a time $t_{\rm obs}$ after the jets
are launched, and the jets proceed at velocity $v_h$,
then $t_b(x)=t_{\rm obs}-x/v_h$. 
 Substituting this in equation (8), keeping the other variables the
same, gives 
\begin{eqnarray} 
z_b \simeq 1.3 \times 10^{16} \left[ \frac {t_{\rm obs}-(x/v_h)}{100 \yr} \right]^{1/2}  
\left( \frac {x} {10^{16} \cm } \right)^{1/2}.
\end{eqnarray} 
 The maximum width of the lobe, $Z_{b {\rm \max}}$, occurs at
\begin{eqnarray} 
x (z_{b{\rm max}}) \simeq \frac {v_h t_{\rm obs}}{2} = 
1.6 \times 10^{17} \cm 
\left( \frac {t_{\rm obs}}{1000 \yr} \right)  
\left( \frac {{\it v_h}} {100 \km \s^{-1}} \right) \cm,
\end{eqnarray}    
 assuming the jet life time was $> t_{\rm obs}/2$.
  For this time of observation, $v_h \simeq 100 \km \s^{-1}$, and with the 
other values in equation (8), the maximum width is $z/x \simeq 0.7$.
 Due to the decrease of the density with distance from the symmetry axis 
and the radial expansion of the slow wind, the lobe will actually be somewhat
wider. 
 These values for the location of the widest lobal point and
its width there are typical for many young bipolar PNe. 
 The derivation above shows that the bubble, or lobe, is narrow close to 
the central system, then becomes wider, and then narrows again. 
 If the density of the material on the outer portion is too low to be 
observed, the observed lobe surface will only widen with distance from 
the equatorial plane.  

 There are several strong assumptions in the derivation above;  
in particular, $(i)$ neglecting the variation of the slow wind
density perpendicular to the symmetry axis; $(ii)$ treating the slow wind
as a static medium rather than a radially expanding medium; and
$(iii)$ the assumption that the bubble surface only expands perpendicular
to the symmetry axis (the $z$ direction). 
 The last assumption is reasonable at the intermediate values of $x$,
but not close to the equatorial plane (small values of $x$).  
 The expansion there is more appropriately described by equation (3) and (4).
 Also, from equation (6) it is clear that the center of each bubble will be
at a large distance from the equator. 
 Hence, the expanding bubble surface facing the equatorial plane will
push material toward the equatorial plane, substantially increasing the 
density there. 
 This supports the claim of SR00 that jets 
(or a CFW) can compress the gas near the equatorial plane, contrary
to some claims that the dense equatorial gas collimates the jets. 
 If the bubble surface reaches the equatorial plane, it will collide
with the bubble on the other side of the equatorial plane, and the
lobe sides will be observed to emerge from the equatorial plane at
a large distance from the center. 
 Such a structure is observed, for example, in the PN He 2-104 
(Sahai \& Trauger 1998).
 The inflation of the bubble away from the central system
implies that the bubble boundary may not reach the equatorial plane.
This effect is stronger if we consider the outward motion of 
the slow wind, which was ignored in the derivation above. 
 However, the distance of the bubble surface from the equatorial plane
will be very small, since the bubble is starting to be inflated by
the jet (or CFW) at $r \sim 400 \AU$ (eq. 6). 
 To observe a bubble not touching the equatorial plane,
the nebula must be observed almost edge on. 
These two constraints imply that only a very small number of such 
bipolar PNe should be observed. 
 Taking into account its inclination, it seem that the lobes of the 
bipolar-PN Hb 12 do not cross the equatorial plane (Sahai \& Trauger 1998).     
 At larger values of $x$ the jets may cease to exist, and a different
treatment is required. 
 Despite these approximations, the treatment above gives the basic
shape of lobes in some bipolar-PNe, and demonstrates the capability
of the binary-blown jets to form bipolar lobes. 

The scaling of the jet velocity, $400 \km \s^{-1}$, is typical for
jets blown by main sequence stars, which have similar escape
velocities (Livio 2000). 
Jets blown by white dwarfs will have much higher velocities, 
of the order of the escape velocity from white 
dwarfs ($\sim 5000 \km \s^{-1}$). 
 Such velocities are directly observed in some PNe and proto-PNe,
e.g., a wind velocity of $2300 \km \s^{-1}$ was reported by 
S\'anchez Contreras \& Sahai (2001) in the proto-PNe He 3-1475.  
 In that case, we find from equation (6) that the shocked jet material 
will not cool, and the bubble will be inflated as soon as the jet
encounters the slow wind gas. 
 In these cases, the lobes that are formed will have a wide opening 
very close to the equatorial plane, e.g., as in the symbiotic Mira 
He 2-104 (the Southern Crab; Corradi {\it et al.} 2001). 

\section{FAST-PROPAGATING JETS}
\subsection{Jet Interaction with the Slow Wind}

 When the density of the gas in the jet is much larger than that
 of the slow wind it will not slow much, and equation (1) is not applicable. 
 Instead, the jet head proceeds at a velocity given by 
\begin{eqnarray} 
v_h \simeq v_j \left[1-\left( \frac{\rho_s}{\rho_j} \right)^{1/2} \right];
\qquad {\rm for} \qquad  \rho_j \gg \rho_s.   
\end{eqnarray} 
 Most of the thermal energy released is that of the shocked slow wind
material, rather than that of the shocked jet material as in the
slow-propagating jet case.  
 Thermal energy injected to the shocked slow wind material per unit
time and unit length are $\dot E_0 = 0.5 A_j \rho_s v_j^3$, and 
\begin{eqnarray}
e_0 = \dot E_0 /v_h \simeq 0.5 A_j \rho_s v_j^2, 
\end{eqnarray}
respectively, where $A_j$ is the cross section of the jet. 

 As in the slow-propagating jet, bubble inflation starts when the shocked
 material has no time to cool. 
  To find this distance $r_s$, the calculations that led to equation (6)
are repeated with the following changes: 
($i$) the energy injection rate $0.5 \dot M_j v_j^2$ in equation (1) is 
replaced by  $\dot E_0$ given above with $A_j = 4 \pi \beta r^2$, and 
($ii$) the shocked material now is the slow wind gas.
 Hence the term for the total shocked mass $\dot M_j t_b$ in the 
derivation of equation (5) is replaced by the mass of the
shocked slow wind during time $t_b$: $\beta \dot M_s t_b v_j/v_s$,  
where, as before, $t_b$ is the expansion time of the bubble. 
 The condition that the cooling time be longer than the expansion time $t_b$,
becomes a condition on the location of the jet head
\begin{eqnarray} 
r_s \gtrsim 6.1 \times 10^{15}
\left( \frac {v_j} {400 \km \s^{-1}} \right)^{-7/3} 
\left( \frac {v_s} {15 \km \s^{-1}} \right)^{-5/6} 
\left( \frac {\dot M_s} {10^{-5} M_\odot \yr ^{-1}} \right) ^{5/6}  
\nonumber \\ \times
\left( \frac {t_b} {100 \yr } \right)^{1/6} 
\left( \frac {\beta} {10^{-3} } \right)^{1/3} \cm .  
\end{eqnarray}
 This condition is very similar to the one for slow-propagating jets
(eq. 6) .
 As with slow-propagating jets, jets blown by main sequence stars,
for which $v_j \simeq 400 \km \s^{-1}$, will start to inflate a bubble
only after propagating a significant distance from their origin; this will
push slow wind material toward the equatorial plane.
 Jets blown by white dwarfs companions, for which 
$v_j \gg 1000 \km \s^{-1}$, on the other hand,  will start to
inflate a bubble very close to their origin at $r \sim 10 \AU$.
 They will also push material toward the equatorial plane, but
in a complicated flow which required 3D numerical simulations
to be treated correctly (SR00).  

\subsection{The Bubble Shape}
  
 I consider two cases which bound the two extremes of a jet's possible 
cross section.  
In the first case, the jet expands radially so that $A_j=4 \pi \beta r^2$, 
at all times.
 In the second case, the jet expands radially at early times, but it 
is then recollimated by the pressure of the surrounding shocked gas or
by an internal magnetic field (e.g., K\"ossl, M\"uller, \& Hillebrandt 1990),
such that its cross section stays constant at $A_j = \pi a_j^2$,
where $a_j$ is the radius of the jet.  
 I do not study these processes here, and simply assume that after 
opening up, the jet is reconfined near the place where bubble 
inflation starts, $r_s \simeq 10^{16} \cm$ by equation (14), and I 
scale the jet radius by $a_j = 0.1 r_s \simeq 10^{15} \cm$. 
 Jets propagating with their cross section increasing at a rate much slower
than $A_j \propto r^2$, and even with constant cross section along
a fraction of their path, are observed in different objects. 
 Examples are the radio jet in the radio galaxy 3C 111 
(Linfield \& Perley 1984) and the molecular outflows in the young 
stellar objects HH 288 (Gueth, Schilke, \& McCaughrean 2001) and 
HH 34 (Reipurth {\it et al.} 2000). 

\subsubsection{Freely expanding jets}

 Substituting the relevant cross section, $A_j = 4 \pi \beta r^2$,
in the rate of thermal energy released per unit length (eq. 13), 
and then in equation (7), gives the radius of the axisymmetric bubble 
as function of the distance, $x$, from the equatorial plane 
\begin{eqnarray} 
\frac {z_b}{x} \simeq 1.0 
\left( \frac {v_j} {400 \km \s^{-1}} \right)^{1/2} 
\left[ \frac {t_b(x)} {100 \yr } \right]^{1/2} 
\left( \frac {x} {10^{16} \cm } \right)^{-1/2}
\left( \frac {\beta} {10^{10^{-3}}} \right)^{1/4}.
\end{eqnarray}
 We find that the basic shape of the lobe is the same as in the case of
a slow-propagating jet, hence, the analysis from equation (8)
to the end of the previous section is applicable to the present
case as well. 

\subsubsection{Reconfined jets}

 Substituting the constant cross section in the thermal energy 
released per unit length (eq. 13), and then in equation (7), gives the
radius of the axisymmetric bubble formed by a reconfined jet 
\begin{eqnarray} 
z_b \simeq 1.3 \times 10^{16}  
\left( \frac {v_j} {400 \km \s^{-1}} \right)^{1/2} 
\left( \frac {a_j} {10^{15} \cm } \right)^{1/2}
\left[ \frac {t_b(x)} {100 \yr } \right]^{1/2} \cm.
\end{eqnarray}
  The shape here is different from that in previous cases; 
rather than widening, the cylindrical lobe has a constant radius. 
 Since the jet propagates outward, the expansion time at location
$x$ from the equatorial plane is $t_{\rm obs}-x/v_j$,
where $t_{\rm obs}$ is the observing time measured from 
the birth of the jet .
Since in the present case the jet propagate much faster than the
slow-propagating case, this delay is less significant than the
case analyzed in equations (10)-(11). 
 In any case, for the same parameters as in the last equation, the 
bubble radius will be 
\begin{eqnarray} 
z_b \simeq 4.0 \times 10^{16}  
\left( \frac {v_j} {400 \km \s^{-1}} \right)^{1/2} 
\left( \frac {a_j} {10^{15} \cm } \right)^{1/2}
\nonumber \\ \times
\left[ \left( \frac {t_{\rm obs}} {1000 \yr } \right) - 
0.40 \left( \frac{x}{5 \times 10^{17} \cm} \right)    
 \left(\frac{v_j}{400 \km \s^{-1}} \right)^{-1}    
\right]^{1/2} \cm.
\end{eqnarray}

 The proto-PN Hen 401 has cylindrical lobes with almost constant
radius of $z_b \simeq 5 \times 10^{16} \cm$ up to a distance of 
$\sim 3.5 \times 10^{17} \cm$ from the central 
source (Sahai, Bujarrabal, \& Zijlstra 1999).
Farther out the lobes become narrower.
 Sahai {\it et al.} (1999) present detailed study of the lobes.
 They argue, based on the theoretical study of molecular outflows
by Masson \& Chernin (1993), that the structure of Hen 401 can
be explained by interaction between a well collimated outflow 
and the slow wind, which intermediate between 
being radiative and being adiabatic. 
Some basic ingredients of their scenario apply here as well. 
 According to the model presented here, the initial jet interaction 
with the slow wind is radiative, i.e., the shocked slow wind material 
has time to cool radiatively.
This is appropriate for a jet blown by a main sequence companion;
the case of a white dwarf companion is discussed below.
 As the jet expands to $r \sim 10^{16} \cm$ (eq. 14 above) the adiabatic
phase starts, with the inflation of a bubble (which will become the observed 
lobes; one lobe at each side of the equatorial plane).
 This part of the bubble compresses the material near the equatorial 
plane; this explains the wide opening of the lobe close to the equatorial
plane, which cannot be explained with a fully radiative flow.
 The jets then expand with a constant cross section.
 Further out, the jets may become weaker, and the interaction becomes
unstable (Sahai {\it et al.} 1999). 
 To quantitatively match the structure of Hen 401, 
$z_b=5 \times 10^{16} \cm$ up to $x=3.5 \times 10^{17} \cm$, 
the age of the jet and its velocity should be known. 
 This information is not available, hence, I consider a few examples.
 We can take $a_j = 3 \times 10^{15} \cm$, keeping the 
other parameters as in equation (17); or we can take 
$t_{\rm obs}=1500 \yr$, $v_j=500 \km \s^{-1}$, and keeping $a_j$ 
as in equation (17).
 For a white dwarf companion, we set $v_j \simeq 5000 \km \s^{-1}$ 
and take $a_j \simeq 2 \times 10^{14} \cm \simeq 13 \AU$, 
keeping the observing time as in equation (17). 
 In this case, the bubble inflation starts much closer to 
the binary system (eq. 14). 
 To summarize, the flow structure presented here, despite its
several assumptions, e.g., neglecting the decrease in the slow wind 
density along the $z$ direction (perpendicular to the symmetry axis), 
can explain the structure of the lobes in Hen 401, and similar PNe 
having cylindrical lobes (see Sahai {\it et al.} 1999
for more examples).
 
\section{SUMMARY}

 In the present paper, I studied the interaction of jets, or a 
collimated fast wind (CFW), with a previously blown AGB wind, and 
reproduced some basic properties of bipolar PN lobes.
 In reality, the jets (or CFW) are blown by an accreting compact companion
simultaneously with the slow wind. 
 To facilitate analytical solution, I had to assume axisymmetric flow,
as well as several other simplifying assumptions.
 In addition, I studied only two extreme cases, a slow-propagating 
and fast-propagating jets, which occur when the slow wind density is much 
larger, or much lower, than the jet density, respectively.   
 The goal was to derive simple expressions from which the properties 
of PN lobes blown by jets can be inferred.
 This can be used both to analyze observations (see discussions following
eqs. 6 and 17), and to guide future numerical simulations. 
 
The main results can be summarized as follows:
\newline
(1) The interaction of jets (or CFW) blown by main sequence stars, 
having a speed of $v_j \sim 300-500 \km \s^{-1}$, with the slow AGB 
stellar wind close to the central binary system is radiative, i.e., 
the shocked gas cools on a short time. 
 Further away, the interaction becomes adiabatic, i.e., a bubble of hot 
gas which cools slowly is formed. 
 The distance from the center where this transition takes place for
the two types of jets studied here is given by equations (6) and (14).
 The expressions are scaled for jets blown by main sequence companions.
Jets blown by white dwarf companions move much faster, 
$v_j \sim 5000 \km \s^{-1}$, and start to inflate a bubble very close
to the binary system. 
\newline
(2) As noted by SR00, the interaction between the CFW and the slow wind
pushes material toward the equatorial plane. 
 The formation of the inflated bubble at a distance from the equatorial 
plane in jets blown by main sequence stars, increases the efficiency of 
this process.
\newline
(3) The significance of the transition from the radiative to the adiabatic
phase to the structure of the lobes, implies that radiative cooling must 
be incorporated in numerical codes intended to study the formation 
of bipolar PNe. 
\newline
(4)  The basic interaction of a radially expanding jet with the 
slow wind forms an axisymmetric bubble which is narrow close 
to the central star(s) and far from the central star(s); its widest 
diameter is somewhere in-between. 
This type of structure is observed in many bipolar PNe. 
\newline
(5) If a jet is recollimated, e.g., by magnetic fields inside the jet,
such that its cross section stays constant, then the lobe widens close 
to the center, but then acquires a cylindrical shape with a constant 
diameter. Such lobes are observed in several bipolar PNe. 
 
\acknowledgements
I thank Liz Blanton for comments on the original manuscript.
This research was supported in part by grants from the 
US-Israel Binational Science Foundation.

\end{document}